\documentclass[english,conference]{IEEEtran}

\usepackage{amsmath,amssymb,bbm,stmaryrd}
\usepackage{bm}
\usepackage{tikz}
\usepackage{stfloats}
\usetikzlibrary{shapes,decorations,matrix,positioning}
\usepackage{pgfplots}
\usepackage{babel,flushend}
\usepackage{scalefnt}
\usepackage{enumerate}
\usepackage[stable]{footmisc}

\newif\ifmaxwell
\maxwelltrue

\newif\iffppotential
\fppotentialtrue

\newif\iftrialPxI
\trialPxIfalse

\newcommand{\R}{\mathbb{R}}
\renewcommand{\t}{\top}
\renewcommand{\epsilon}{\varepsilon}

\newcommand{\x}{\bm{x}}
\newcommand{\y}{\bm{y}}
\newcommand{\z}{\bm{z}}
\newcommand{\A}{\bm{A}}

\newcommand{\D}{\bm{D}}
\newcommand{\f}{\bm{f}}
\newcommand{\g}{\bm{g}}
\renewcommand{\vec}[1]{\text{vec}\!\left(#1\right)}

\newcommand{\e}{\bm{\epsilon}}
\providecommand{\norm}[1]{\left\lVert#1\right\rVert}

\renewcommand{\S}{\bm{S}}
\newcommand{\X}{\bm{X}}
\newcommand{\Xc}{\mathcal{X}}
\newcommand{\Xm}{\mathcal{X}_{\circ}}
\newcommand{\Ec}{\mathcal{E}}
\newcommand{\Em}{\mathcal{E}_{\circ}}

\newcommand{\Z}{\bm{Z}}

\newtheorem{theorem}{Theorem}
\newtheorem{lemma}{Lemma}
\newtheorem{definition}{Definition}

\newtheorem{remark}{Remark}
\newtheorem{corollary}{Corollary}

\pgfplotsset{tick label style={
font=\Large}}

\IEEEoverridecommandlockouts

\begin{document}
\pgfdeclarelayer{background}
\pgfdeclarelayer{foreground}
\pgfsetlayers{background,main,foreground}

\title{A Simple Proof of Threshold Saturation \\ for Coupled Vector Recursions}

\author{Arvind Yedla, Yung-Yih Jian, Phong S. Nguyen, and Henry D. Pfister%
\thanks{This material is based upon work supported in part by the
  National Science Foundation (NSF) under Grants No. 0747470.
  Any opinions, findings, conclusions, and recommendations expressed
  in this material are those of the authors and do not necessarily
  reflect the views of the NSF.}
\\Department of Electrical and Computer Engineering, Texas A\&M University}

\maketitle

\begin{abstract}
  Convolutional low-density parity-check (LDPC) codes (or
  spatially-coupled codes) have now been shown to achieve capacity on
  binary-input memoryless symmetric channels.
  The principle behind this surprising result is the threshold-saturation phenomenon,
  which is defined by the belief-propagation threshold of the
  spatially-coupled ensemble improving to a fundamental threshold defined
  by the uncoupled system.

  Previously, the authors demonstrated that potential functions
  can be used to provide a simple proof of threshold saturation for
  coupled scalar recursions.  In this paper, we present a simple proof
  of threshold saturation that applies to a wide class of coupled
  vector recursions.  The conditions of the theorem are verified for
  the density-evolution equations of: (i) joint decoding of irregular
  LDPC codes for a Slepian-Wolf problem with erasures, (ii) joint decoding of
  irregular LDPC codes on an erasure multiple-access channel, and (iii) protograph codes on the BEC.
  This proves threshold saturation for these systems.
\end{abstract}

\section{Introduction}

Low-density parity-check (LDPC) convolutional codes, or
spatially-coupled (SC) LDPC codes, were introduced in
\cite{Felstrom-it99} and observed to have excellent belief-propagation
(BP) thresholds in
\cite{Sridharan-aller04,Lentmaier-isit05,Lentmaier-it10}.  Recently,
they have been observed to approach capacity for a variety of
problems~\cite{Lentmaier-it10,Kudekar-istc10,Rathi-isit11,Yedla-isit11,Kudekar-isit11-DEC,Nguyen-arxiv11,Nguyen-icc12,Kudekar-arxiv12}.

The principle behind their excellent performance is
described in \cite{Kudekar-it11}, where it is shown analytically for
the BEC that the BP threshold of a regular SC ensemble converges to
the maximum-a-posteriori (MAP) threshold of the uncoupled ensemble.
This phenomenon is now called \emph{threshold saturation}.  A similar
observation was reported independently in \cite{Lentmaier-isit10} and
stated as a conjecture.  For binary-input memoryless symmetric
(BMS) channels, threshold saturation was empirically observed
first~\cite{Lentmaier-it10,Kudekar-istc10} and then shown
analytically~\cite{Kudekar-arxiv12}.

Threshold saturation now appears to be quite general and spatial-coupling
has now been applied, with great success, to
more general scenarios in information theory and coding%
~\cite{Hassani-itw10,Hassani-jsm12,Takeuchi-isit11,Schlegel-isit11,Jian-isit12,Kudekar-aller10,Krzakala-arxiv11,Donoho-arxiv11}

Recently, the authors introduced a simple proof of threshold saturation
for coupled scalar recursions where only a few details must be
verified for each system~\cite{Yedla-istc12}.
The examples presented therein prove a number of threshold saturation
conjectures made in the aforementioned
papers~(e.g., see~\cite{Kudekar-isit11-DEC,Nguyen-icc12}).
The proof technique is based on potential functions for
density-evolution (DE) recursions and was motivated by the ideas
in~\cite{Takeuchi-arxiv11}. Another approach to proving threshold
saturation, based on a continuous approximation of spatial-coupling,
can be found in~\cite{Donoho-arxiv11} and~\cite{Kudekar-unpub12}.

In this paper, the analysis is extended to prove that threshold
saturation also occurs for certain coupled systems of vector
recursions.  In particular, if the single-system vector recursion can
be generated by a scalar potential function, then one can also define
a scalar potential function for the coupled system.  Using this, one
can show that threshold saturation occurs for a wider class of
problems.  For example, this settles conjectures made
in~\cite{Yedla-isit11,Kudekar-isit11-MAC}.  Along with the results
in~\cite{Nguyen-isit12}, this shows the universality of SC codes for
a noisy Slepian-Wolf problem with erasures, when the channels are unknown
at the transmitter.


\section{A Simple Proof of Threshold Saturation}
\label{sec:PotentialThm}

In this section, we provide a simple proof of threshold saturation via
spatial-coupling for a broad class of vector recursions.  The main
tool is a potential theory for vector recursions that extends
naturally to coupled systems.

\subsection{Notation}
\label{sec:notation}
The following notation is used throughout this paper.
We let $d\in\mathbb{N}$ be the dimension for the vector recursion, $\Xc \triangleq [0,1]^d$ be the space on which the recursion is defined, and $\Ec \triangleq [0,1]$ be the parameter space of the recursive system.
For convenience, we let $\Xm \triangleq \Xc\setminus\{\bm{0}\}$ and $\Em \triangleq \Ec\setminus\{0\}$.
Vectors are denoted in boldface lowercase (e.g. $\x,\y$), are assumed to be row vectors, and inherit the natural partial order $\x\preceq \y$ defined by $x_i\leq y_i$ for $1 \le i \le d$.
Matrices are denoted in boldface capital letters (e.g. $\X \in \Xc^n$) and we use $\x_i=[\X]_i$ to denote the $i$-th row of $\X$ and $x_{i,j} = [\X]_{i,j}$ to denote the $(i,j)$-th element of $\X$.
Standard-weight typeface is used for scalar-valued functions with a vector/matrix argument (e.g., $F(\x)$,$F(\X)$) and boldface is used to denote a vector-valued function with a vector argument (e.g., $\f(\x) = [f_1(\x),\cdots, f_d(\x)]$).
The gradient of a scalar function is defined by $F'(\x) \triangleq [\partial F(\x)/\partial x_1,\cdots, \partial F(\x)/\partial x_d]$ and the Jacobian of a vector function is defined by
\vspace{-2mm}
\begin{align*}
  \f'(\x) = \frac{\partial \f(\x)}{\partial \x} \triangleq
  \begin{bmatrix}
    \frac{\partial f_1(\x)}{\partial x_1} & \cdots & \frac{\partial
      f_1(\x)}{\partial x_d} \\
    \vdots & \ddots & \vdots \\
    \frac{\partial f_d(\x)}{\partial x_1} & \cdots & \frac{\partial
     f_d(\x)}{\partial x_d}
  \end{bmatrix}.
\end{align*}
The gradient of a scalar-valued function with a matrix argument
denoted by $F'(\X)$ is a matrix defined by $[F'(\X)]_{i,j}
\triangleq
\partial F(\X)/\partial x_{i,j}$. For two-argument functionals
(e.g., $F(\x;\epsilon)$ and $U(\x;\epsilon)$), we define
$F'(\x;\epsilon) \triangleq \partial F(\x;\epsilon)/\partial \x$.

Abusing notation, we also allow functions defined for vector arguments (e.g., $\f(\x)$) to act on matrices (e.g., $\f(\X)$) via the rule $[\f(\X)]_i = \f(\x_i)$.
The notation $\vec{\X}$ denotes the transpose of the vector obtained by stacking the columns of $\X$~\cite{Magnus-1999}. 
The Jacobian and Hessian of a matrix function are denoted by
\begin{align*}
  \f'(\X) \triangleq \frac{\partial\vec{\f(\X)}}{\partial\vec{\X}}
  \quad \text{and} \quad
  \f''(\x) \triangleq \frac{\partial \vec{\f'(\x)}}{\partial\x}.
\end{align*}

\subsection{Single-System Potential}
\label{sec:single-sytem-potent}

First, we define potential functions for a class of vector
recursions and discuss the associated thresholds.

\begin{definition}
\label{def:vas}
Let $F : \Xc \times \Ec \to \R$ and $G : \Xc \to \R$ be functionals
and $\D$ be a $d \times d$ positive diagonal matrix.
Consider the recursion defined by
  \begin{align}
    \label{eq:1}
    \x^{(\ell+1)} = \f(\g(\x^{(\ell)});\epsilon),
  \end{align}
where $\f : \Xc \times \Ec \to \Xc$ and $\g : \Xc \to \Xc$ are
mappings defined by $F'(\x;\epsilon) = \f(\x;\epsilon) \D$
and $G'(\x) = \g(\x) \D$.
Then, the pair~$(\f,\g)$ defines a \emph{vector admissible system} if
\begin{enumerate}[i)]
\item $\f,\g$ are twice continuously differentiable,
\item $\f(\x;\epsilon),\g(\x)$ are non-decreasing in $\x$ (w.r.t. $\preceq$),
\item $\f(\x;\epsilon)$ is strictly increasing in $\epsilon$ for $\x \in \Xm$,
\item $\f(\bm{0};\epsilon) = \f(\x;0) = \g(\bm{0}) = \bm{0}$ and $F(\x;0)=0$.
\end{enumerate}
\end{definition}

\begin{remark}
\label{rem:vecpara}

More generally, the vector recursion may have a parameter vector
$\e \in \Ec^n$ and be defined by
\begin{align}
 \label{eq:vecparam}
 \x^{(\ell+1)} & = \widetilde{\f}(\g(\x^{(\ell)});\e).
\end{align}
In this case, one can consider a path $\e(\epsilon) \in \Ec^n$
parameterized by $\epsilon \in \Ec$. Such a path is called
valid if it is smooth, strictly increasing (w.r.t. the
partial order) in $\epsilon$, $\e(0)={\bf 0}$, and $\e(1)={\bf 1}$.
The recursion in~\eqref{eq:vecparam} along a valid path can
be converted into a new recursion in the form of~\eqref{eq:1} with
\[ \x^{(\ell+1)} =  \f(\g(\x^{(\ell)});\epsilon) \triangleq \widetilde{\f}(\g(\x^{(\ell)});\e(\epsilon)). \]
If the resulting functions $(\f, \g)$ satisfy the conditions in
Def.~\ref{def:vas} for any valid $\e(\epsilon)$ path, then the
recursion in~\eqref{eq:vecparam} can be characterized using a scalar
$\epsilon$ analysis and a family of $\e(\epsilon)$ paths.
This is the approach taken in this work.
\end{remark}

\begin{lemma}
\label{lem:LineSegment}
For any $(\x,\epsilon) \in \Xc\times\Ec$ and $t\in [0,1]$, we
define $\z(t) \triangleq \x+t(\f(\g(\x);\epsilon)-\x)$.
If $\x \preceq \f(\g(\x);\epsilon)$, then $\z(t) \preceq
\f(\g(\z(t));\epsilon)$ for all $t\in[0,1]$. Similarly, if
$\x \succeq \f(\g(\x);\epsilon)$, then $\z(t) \succeq
\f(\g(\z(t));\epsilon)$ for all $t\in[0,1]$.
\end{lemma}
\begin{IEEEproof}
First, consider the case where $\x \preceq \f(\g(\x);\epsilon)$.
It is easy to verify that $\x \preceq \z(t) \preceq \f(\g(\x);\epsilon)$.
Since $\f$ and $\g$ are non-decreasing, one can also observe that
\begin{equation*}
 \z(t) \preceq \f(\g(\x);\epsilon) \preceq \f(\g(\z(t));\epsilon).
\end{equation*}
Thus, the first claim follows.
If $\x \succeq \f(\g(\x);\epsilon)$, the same approach shows that $\z(t) \succeq \f(\g(\x);\epsilon) \succeq \f(\g(\z(t));\epsilon)$.
\end{IEEEproof}

\begin{definition}
  \label{def:potential}
  Let the \emph{potential function} $U(\x;\epsilon)$
of a vector admissible system~$(\f,\g)$ be defined by
  \begin{align}
    \label{eq:2}
    U(\x;\epsilon) &\triangleq \int_0^{\x}\left[\left(\z - \f(\g(\z);\epsilon)\right)\D \g'(\z)\right]\cdot\mathrm{d}\z \notag\\
    &= \g(\x) \D \x^\t - G(\x) - F(\g(\x);\epsilon),
  \end{align}
where $U(\bm{0};\epsilon)=0$ implies $F(\bm{0};\epsilon)=G(\bm{0})=0$.
\end{definition}

\begin{remark}
The above result is easily verified by taking the derivative of~\eqref{eq:2} with respect to $\x$.
Of course, this also implies that the line integral is path independent.
For DE equations, the potential function is related to the pseudo-dual of the
Bethe free energy (e.g., see~\cite[Part 2, pp.~62-65]{Vontobel-acorn09} \cite{Walsh-com10}).

\end{remark}

\begin{lemma}
\label{lem:ReducePotential}
For any $(\x,\epsilon) \in \Xc\times\Ec$, let $\y \triangleq
\f(\g(\x);\epsilon)$. \linebreak If $\x \preceq \y$ or $\x
\succeq \y$, then $U(\x;\epsilon) \geq U(\y;\epsilon)$.
\end{lemma}
\begin{IEEEproof}
Consider the path from $\x$ to $\y$ defined by \linebreak $\z(t) \triangleq \x+t(\y-\x)$ for $t\in[0,1]$.
By Def.~\ref{def:potential}, one can write
\begin{align}
 \label{eq:2step}
 U\left(\y;\epsilon \right) & = U\left(\x;\epsilon \right) \notag \\
 & + \!\! \int^{1}_{0} \!\!\! \left[(\z(t) \! - \!\! \f(\g
 (\z(t));\epsilon))\D\g'\!(\z(t))\right]\! \cdot \! \z'(t) \mathrm{d}t.\!\!\!
\end{align}
If $\x \preceq \y$, then Lem.~\ref{lem:LineSegment} implies that $\z(t)-\f(\g(\z(t));\epsilon) \preceq \bm{0}$
and $\z'(t) \succeq \bm{0}$ for all $t\in[0,1]$.
Next, we note that $\g(\x)$ non-decreasing w.r.t. $\preceq$ implies that $\g'(\x)$ is a non-negative matrix.
Since $\D$ and $\g'(\x)$ are non-negative matrices and $\z'(t) \succeq \bm{0}$ for $t\in[0,1]$, the integral in (\ref{eq:2step}) is upper bounded by $0$.
Similarly, if $\x \succeq \y$, then $\z'(t) \preceq \bm{0}$ and $\z(t)-\f(\g(\z(t));\epsilon)\succeq \bm{0}$.
Again, this implies the integral in (\ref{eq:2step}) is upper bounded by $0$.
Therefore, we conclude that $U\left( \y;\epsilon \right) \leq U\left( \x;\epsilon \right)$.
\end{IEEEproof}

\begin{definition}
\label{def:fp_sp}
For $\x\in\Xc$ and $\epsilon\in\Ec$,
\begin{enumerate}[i)]
\item
 $\x$ is a \emph{fixed point} (f.p.) if $\x = \f(\g(\x);\epsilon)$.
\item
 $\x$ is a \emph{stationary point} (s.p.) if $U'(\x;\epsilon) = \bm{0}$.
\end{enumerate}
\end{definition}

\ifmaxwell
\begin{definition}
\label{def:eroot}
The \emph{fixed-point set}, $\mathcal{F}(\epsilon)$, its $\x$-support, $\mathcal{X}_{f}$, and the \emph{epsilon set}, $\epsilon(\x)$, are given by
\begin{align*}
  \mathcal{F}(\epsilon) &\triangleq \left\{\x \in \Xm \mid \x = \f(\g(\x);\epsilon)\right\}\\
  \mathcal{X}_f &\triangleq \cup_{\epsilon \in \Ec} \mathcal{F}(\epsilon) \\
  \epsilon(\x) &\triangleq \{ \epsilon \in \Ec \mid \x \in \mathcal{F}(\epsilon) \}.
\end{align*}
\end{definition}
\fi

\begin{lemma}
\label{lem:3properties}
For a vector admissible system, we have
  \begin{enumerate}[i)]
  \item $U(\x;\epsilon)$ is strictly decreasing in $\epsilon$, for $\x \in \Xm$ and
    $\epsilon \in \Ec$.
  \item Any f.p. $\x\in\Xc$ is a s.p. of the potential.
\ifmaxwell
  \item For any $x \in \mathcal{X}_{f}$, $\epsilon(\x)$ has a single element.
\fi
  \end{enumerate}
\end{lemma}
\begin{IEEEproof}
 The potential function is a line integral of $\left(\z - \f(\g(\z);\epsilon)\right) \D \g'(\z)$, and this expression is strictly decreasing in $\epsilon$, for $\x \in \Xm$ and $\epsilon \in \Ec$.
   The second property also follows from this formulation because, at a f.p., the integrand is $\bm{0}$.
\ifmaxwell
  For the third, let $\x \in \mathcal{X}_{f}$ and observe that the definition of $\mathcal{X}_{f}$ implies $\epsilon(\x)$ has at least one element.
  Since $\f(\x;\epsilon)$ is strictly increasing in $\epsilon$, for $\x \in \Xm$ and $\epsilon\in\Ec$, it follows that $\x = \f(\g(\x);\epsilon)$ can have at most one $\epsilon$-root.
  Therefore, $\epsilon(\x)$ is a singleton and we treat it as a function $\epsilon: \mathcal{X}_f \to \Ec$.
\fi
\end{IEEEproof}

\begin{definition}
  For a pair $(\x,\epsilon)\in \Xc \times \Ec$, let $\x^{(0)} = \x$ be the initial value of the recursion defined in (\ref{eq:1}).
  We define $\x^{\infty}(\x;\epsilon) = \lim_{\ell \rightarrow \infty}  \f(\g(\x^{(\ell)});\epsilon)$ if the limit exits.
\end{definition}

\begin{lemma}
\label{lem:LimitExists}
For any $(\x,\epsilon) \in \Xc\times\Ec$, if $\x \preceq \f(\g(\x);\epsilon)$ or $\x \succeq \f(\g(\x);\epsilon)$,
then $\x^{\infty}(\x;\epsilon)$ exists and
\begin{equation}
\label{eq:FPisLocalMin}
U(\x;\epsilon)\geq U(\x^{\infty}(\x;\epsilon);\epsilon).
\end{equation}
\end{lemma}
\begin{IEEEproof}
Consider the case where $\x \preceq \f(\g(\x);\epsilon)$, and let
$\x^{(0)}=\x$. By the fact that $\x^{(1)}=\f(\g(\x);\epsilon)
\succeq \x^{(0)}$, the sequence of the vectors $\x^{(\ell)}$ for
$\ell \geq 0$ satisfies $\x^{(0)} \preceq \x^{(1)} \preceq \x^{(2)}
\preceq \cdots \preceq \bm{1}$.
This implies that, for $i=1,\ldots,d$, $[\x^{(\ell)}]_i$ is a non-decreasing real sequence upper bounded by 1.
Therefore, the limit $\x^{\infty}(\x;\epsilon)$ exists.
Similarly, if $\x \succeq \f(\g(\x);\epsilon)$, then the limit exists because each $[\x^{(\ell)}]_i$ is non-increasing and lower bounded by 0.
Finally, Lem.~\ref{lem:ReducePotential} implies that, for all $\ell \geq 0$, $U(\x^{(\ell)};\epsilon)\geq U(\x^{(\ell+1)}; \epsilon)$.
Therefore, the inequality (\ref{eq:FPisLocalMin}) follows.
\end{IEEEproof}

\begin{corollary}
For all $\epsilon \in \Ec$, $\x^{\infty}(\bm{1};\epsilon)$ exists.
\end{corollary}
\begin{IEEEproof}
  This follows from $\bm{1} \succeq \f(\g( \bm{1});\epsilon)$ for $\epsilon \in \Ec$.
\end{IEEEproof}

\begin{definition}
  The \emph{single-system threshold} is defined to be
\begin{align*}
  \epsilon_s^* &\triangleq \sup\left\{ \epsilon\in \Ec \mid
    \x^{\infty}(\bm{1};\epsilon) = \bm{0} \right\},
\end{align*}
and is the $\epsilon$-threshold for convergence of the
recursion to $\bm{0}$.
\end{definition}

\begin{remark}
  The recursion (\ref{eq:1}) has no f.p.s in $\Xm$ iff $\epsilon < \epsilon_s^*$.
  For DE recursions associated with BP decoding, the threshold $\epsilon_s^*$ is called the BP threshold.
\end{remark}

\begin{definition}
\label{def:PotentialThreshold}
  The \emph{potential threshold} is defined by
  \begin{equation}
    \label{eq:3}
    \epsilon^{*} \triangleq \sup \{ \epsilon \in \Ec \mid \min\nolimits_{\x \in \mathcal{F}(\epsilon)} U(\x;\epsilon) \geq 0 \}.
  \end{equation}
  This is well defined because $U(\x;0)\geq 0$.
  For $\epsilon > \epsilon_s^*$, the quantity $\Delta E(\epsilon) \triangleq \min_{ \x\in \mathcal{F}(\epsilon)} U(\x;\epsilon)$ is the called \emph{energy gap}
  and $\epsilon_s^* < \epsilon < \epsilon^{*}$ implies $\Delta E(\epsilon)>0$.
  Since $U(\x;\epsilon)$ is strictly decreasing in $\epsilon$, $\epsilon^* \leq \epsilon'$ for any $\epsilon'$ that satisfies $\Delta E(\epsilon')=0$.
\end{definition}

\begin{definition}
 \label{def:FixedPointPotential}
 The \emph{fixed-point potential}, $Q: \mathcal{X}_f \to \mathbb{R}$, is defined by $Q(\x) = U(\x;\epsilon(\x))$.
 The potential threshold defined in (\ref{eq:3}) can be rewritten as
 \begin{equation}
   \label{eq:PotentialInQ}
   \epsilon^{*} \triangleq \sup \{ \epsilon \in \Ec \mid \min\nolimits_{\x \in \mathcal{F}(\epsilon)} Q(\x) \geq 0 \}.
 \end{equation}
\end{definition}

\iffppotential
\begin{remark}
\label{rem:FixedPointPotential}
 Consider the fixed points $\x_{1}, \x_{2} \in \Xc_{f}$ and observe that
 $Q(\x_{1}) = U(\x_{1};\epsilon(\x_{1}))$ and $Q(\x_{2}) = U(\x_{2};\epsilon(\x_{2}))$.
 Suppose there is a differentiable path $\z(t) \in \Xc_{f}$ from $\x_{1}$ to $\x_{2}$ such that $\epsilon(\z(t))$ is differentiable for $t\in[0,1]$.
 Then, $Q'(\x)$ exists and
 \begin{align}
 \label{eq:intQ'}
  \!Q(\x_{2}) \! &\!= \! Q(\x_{1}) \! + \!\!\! \int^{\x_{2}}_{\x_{1}} Q'(\z)\cdot \mathrm{d}\z \notag\\
   &\! = \! Q(\x_{1}) \! + \!\!\! \int^{1}_{0} \!\! U^{(0,1)}(\z(t);\epsilon(\z(t))) \epsilon'(\z(t))\! \cdot \!\z'(t)
   \mathrm{d}t.\!\!
 \end{align}

\end{remark}
\else

\ifmaxwell
\begin{definition}[{\cite[Conj.~1]{Measson-it08}}]
\label{def:Maxwell}
 Consider a curve $\x(t)$ in $\Xc$ characterized by a single parameter $t\in[0,1]$ with $\x(0)=\bf{0}$ and $\x(1)=\bf{1}$.
 Let $(\x(t), \e(\x(t)))$ be the fixed points of the DE equations along the curve.
 Suppose that $\e(\x(t))$ is continuous in $t$, we can reparameterize the curve $\e(\x(t))$ by $\epsilon \in \Ec$ with $\e(1)=\bf{1}$.
 The \emph{Maxwell threshold} $\epsilon^{\mathrm{Max}}$ is defined
 by
 \begin{equation}
    \label{eq:maxwell_thresh}
     \epsilon^{\mathrm{Max}} \triangleq \inf \{ \epsilon \in \Ec \mid \max\nolimits_{\x \in \mathcal{F}(\epsilon)} P(\x) > 0 \},
 \end{equation}
 where $P(\x)$ is the \emph{trial entropy} from \cite[Lemma 4]{RU-2008, Nguyen-isit12}.
\end{definition}

\fi
\fi

The following lemma compares the minimum fixed-point potentials of
two different $\epsilon$ parameters.
\begin{lemma}
\label{lem:CompareMins}

Let $\epsilon_1, \epsilon_2 \in \Ec$ satisfy $\epsilon_1 < \epsilon_2$, $\mathcal{F}(\epsilon_1) \neq \emptyset$, and $\mathcal{F}(\epsilon_2) \neq \emptyset$.
Then,
\begin{equation}
 \label{eq:qmin12}
 \min\nolimits_{\x \in \mathcal{F}(\epsilon_1)} Q(\x) > \min\nolimits_{\x \in \mathcal{F}(\epsilon_2)} Q(\x).
\end{equation}

\end{lemma}
\begin{IEEEproof}
Let $\x_1 \in \mathcal{F}(\epsilon_1)$ and $\x_2 \in \mathcal{F}(\epsilon_2)$
be the fixed points that achieve the minimums in~\eqref{eq:qmin12}.
Since $\f(\x;\epsilon)$ is strictly increasing in $\epsilon$, one finds that
\begin{align*}
 \x_1 & = \f(\g(\x_1);\epsilon_1) \preceq \f(\g(\x_1);\epsilon_2) \preceq
 \x^{\infty}(\x_1;\epsilon_2).
\end{align*}
Using $\x^{\infty}(\x_1;\epsilon_2) \in \mathcal{F}(\epsilon_2)$, the proof concludes with
\begin{align*}
 \min_{\x \in \mathcal{F}(\epsilon_1)} \! Q(\x) & = U(\x_1;\epsilon_1) \! \overset{(a)}{>} \! U(\x_1;\epsilon_2)
 \! \overset{(b)}{\geq} \! U(\x^{\infty}(\x_1;\epsilon_2);\epsilon_2) \\
  & \geq \min_{\x \in \mathcal{F}(\epsilon_2)} U(\x;\epsilon_2)= \min_{\x \in \mathcal{F}(\epsilon_2)} Q(\x),
\end{align*}
where the inequality $(a)$ is from Lem.~\ref{lem:3properties}, and the
inequality $(b)$ is the result of Lem.~\ref{lem:LimitExists}.
\end{IEEEproof}

\begin{remark}
\label{rem:StrictlyDecreaseQ}

Since $\min_{\x \in \mathcal{F}(\epsilon)} Q(\x)$ is strictly
decreasing in $\epsilon$, one finds $U(\x;\epsilon)>0$ for all $\x
\in \mathcal{F}(\epsilon)$ when $\epsilon\in(\epsilon^*_s,
\epsilon^*)$. If $\epsilon^* < \epsilon \leq 1$, then one also has
$\min_{\x \in \mathcal{F}(\epsilon)} Q(\x)<0$.
\end{remark}



\section{Coupled-System Potential}
\label{sec:coupl-syst-potent}

We now extend our definition of potential functions to coupled
systems of vector recursions.  In particular, we consider a
``spatial-coupling'' of the single system recursion, (\ref{eq:1}),
that gives rise to the recursion (\ref{eq:4}) and a closely related
matrix recursion (\ref{eq:spatial_fp}).  For the
matrix recursion of the coupled system, we define a potential
function and show that, for $\epsilon < \epsilon^*$, the only fixed
point of the coupled system is the zero matrix.
We note that a similar potential was defined earlier for the special
case of a scalar Curie-Weiss system in~\cite{Hassani-jsm12}.

\begin{definition}[cf.~\cite{Kudekar-it11}]
\label{def:SpatiallyCoupled}
The basic \emph{spatially-coupled vector system} is defined by placing
$2L+1$ $\f$-systems at positions in the set $\mathcal{L}_{\f} =
\{-L,-L+1,\ldots,L\}$ and coupling them with $2L+w$ $\g$-systems at positions in the set $\mathcal{L}_{\g} = \{-L,-L+1,\ldots,L+(w-1)\}$.
  For the coupled system, this leads to the recursion, for $i \in \mathcal{L}_{\g}$, given by
\vspace{-1mm}
\begin{align}
\label{eq:4}
  \x^{(\ell+1)}_i &=
  \frac{1}{w}\sum_{k=0}^{w-1}\f\Bigg(\frac{1}{w}\sum_{j=0}^{w-1}\g(\x^{(\ell)}_{i+j-k});\epsilon_{i-k}\Bigg),
  \vspace{-2mm}
\end{align}
where $\epsilon_{i} = \epsilon$ for $i\in \mathcal{L}_{\f}$,
$\epsilon_i = 0$ for $i\notin \mathcal{L}_{\f}$, $\x_i^{(0)} =
\bm{1}$ for $i\in \mathcal{L}_{\g}$, and $\x^{(\ell)}_i = \bm{0}$
for $i \not \in \mathcal{L}_{\g}$ and all $\ell$.
\end{definition}

\begin{definition}[cf.~\cite{Yedla-istc12}]
  The recursion in~(\ref{eq:4}) can be rewritten as a
  \emph{matrix recursion}.  Let $\X \in \Xc^{2L+w}$ have the decomposition
  $\X = [\x_{-L}^{\t},\cdots,\x_{L+w-1}^{\t}]^{\t}$.
  Then,~(\ref{eq:4}) is given by
\begin{align}
\label{eq:spatial_fp}
  \X^{(\ell+1)} = \A^{\t}\f(\A  \g(\X^{(\ell)});\epsilon),
\end{align}
where $\A$ is the following $(2L+1)\times(2L+w)$ matrix. 
\begin{center}
\hspace{9mm}
\begin{tikzpicture}[decoration=brace]
    \matrix (m) [matrix of math nodes,left delimiter=[,right delimiter={]}] {
        1 & 1 & \cdots & 1 & 0 & 0 & \cdots & 0\\
        0 & 1 & 1 & \cdots & 1 & 0 & \smash{\cdots} & 0 \\
        \vphantom{0}\smash{\vdots} & \smash{\ddots} & \smash{\ddots} & \smash{\ddots} & \smash{\ddots} & \smash{\ddots} & \smash{\ddots} & \smash{\vdots} \\
        0 & \cdots & 0 & 1 & 1 & \cdots & 1 & 0\\
        0 & \cdots & 0 & 0 & 1 & 1 & \cdots & 1\\
    };
    \node[transform canvas={xshift=-3.35cm},thick] (m-2-1.west) {$\A =\dfrac{1}{w}$};
    \draw[decorate,transform canvas={xshift=0.4cm},thick] (m-1-8.north east) -- node[right=2pt] {\rotatebox{270}{\small$2L+1$}} (m-5-8.south east);
    \draw[decorate,transform canvas={yshift=0.0em},thick] (m-1-1.north west) -- node[above=2pt] {\small$w$} (m-1-4.north east);
    \draw[decorate,transform canvas={yshift=-0.0em},thick] (m-5-8.south east) -- node[below=2pt] {\small$2L+w$} (m-5-1.south west);
\end{tikzpicture}
\end{center}
\end{definition}

\begin{definition}[cf.~\cite{Kudekar-it11}]
\label{def:OneSidedSC} Let $i_0 \triangleq \lfloor \frac{w-1}{2}
\rfloor$. The \emph{one-sided spatially-coupled vector system} is a
modification of (\ref{eq:4}) defined by fixing the values of
positions outside $\mathcal{L}_{\f}'=\{-L,\ldots,i_0\}$.
It forces the remaining values to $\x_{i_{0}}^{(\ell)}$, that is $\x_{i}^{(\ell)} = \x_{i_{0}}^{(\ell)}$ for $i_{0} < i \le 2L+(w-1)$ and all $\ell$.
\end{definition}

\begin{lemma}[cf.~{\cite[Lem. 14]{Kudekar-it11}}]
\label{lem:monotone}
For both the basic and one-sided SC systems, the recursions are
componentwise decreasing with iteration and converge to well-defined
fixed points.  The one-sided recursion in Def.~\ref{def:OneSidedSC} is a componentwise upper
bound on the basic SC recursion for $i\in\mathcal{L}_{\g}$ and converges
to a non-decreasing fixed-point vector. 
\end{lemma}
\begin{IEEEproof}[Sketch of Proof]
  The proof follows from the monotonicity of $\f,\g$ and a careful treatment of the boundary conditions.
\end{IEEEproof}

\begin{definition}
\label{def:coupledpotential}
The \emph{coupled-system potential} for admissible matrix recursions
is defined to be
\vspace{-1mm}
\begin{align*}
  U(\X;\epsilon) &\triangleq  \mathrm{Tr}\big(\g(\X) \D \X^\t \big) - G(\X) - F(\A \g(\X);\epsilon),
  \vspace{-1mm}
\end{align*}
where $G(\X) = \sum_i G(\x_i)$ and $F(\X;\epsilon) = \sum_i
F(\x_i;\epsilon)$.
\end{definition}

\begin{remark}
The key property of the above coupled-system potential is that the derivative w.r.t. $[\X]_i = \x_i$ has the form
 \begin{align*}
  [U'(\X;\epsilon)]_i &= \left(\x_i -
    [\A^{\t}\f(\A\g(\X);\epsilon)]_i\right) \D \g'(\x_i).
 \end{align*}
To see this, one can compute the derivative of each term separately.
For the first term, we have
\begin{align*}
\frac{\partial}{\partial \x_i} \mathrm{Tr}\big(\g(\X) \D \X^\t \big)
&= \frac{\partial}{\partial \x_i} \sum_{j=-L}^{L+w-1} \g(\x_{j})\D\x_{j}^{\t} \\
&= \x_{i} \D \g'(\x_{i}) + \g(\x_{i})\D.
\end{align*}
For the second term, we use $ (\partial/\partial \x_i) G(\X) = \g(\x_i) \D$.
For the third term, we use
\begin{multline*}
\frac{\partial}{\partial \x_i} F(\A \g(\X);\epsilon)
= \frac{\partial}{\partial \x_i} \sum_{j=-L}^{L} F\left( \sum_{k=-L}^{L+w-1} a_{j,k} \, \g(\x_{k});\epsilon \right) \\
= \sum_{j=-L}^{L} a_{j,i} \, \f\left( \sum_{k=-L}^{L+w-1} a_{j,k} \, \g(\x_{k});\epsilon \right) \D \g'(\x_i) \\
= [\A^{\t}\f(\A\g(\X);\epsilon)]_i \D \g'(\x_i).
\end{multline*}
\end{remark}

\begin{definition}
The down-shift operator $\S_n:\Xc^{n}\to\Xc^{n}$ is defined by $\left[\S_n \X\right]_{1}=\bm{0}$ and $\left[\S_n \X\right]_{i}=\x_{i-1}$ for $2 \le i \le n$.
In the sequel, the subscript of the down-shift operator is omitted, and it can be inferred from the context.
\end{definition}

\begin{lemma}
\label{lem:shiftnormbound}
Let $\X \in \Xc^n$ be a matrix with non-decreasing columns generated by
averaging the rows of $\Z\in\Xc^n$ over a sliding window of size $w$.
Then, $\norm{\vec{\S\X-\X}}_{\infty} \leq
\frac{1}{w}$ and $\norm{\vec{\S\X-\X}}_1 = \norm{\x_n}_1 = \norm{\X}_{\infty}$.
\end{lemma}
\begin{IEEEproof}
For $\norm{\vec{\S\X-\X}}_{\infty}$, one has
\begin{align*}
\left|x_{i,j}-x_{i-1,j}\right| &= \left|\tfrac{1}{w}
\textstyle{\sum_{k=0}^{w-1}} z_{i+k,j} - \tfrac{1}{w}
\textstyle{\sum_{k=0}^{w-1}} z_{i-1+k,j}\right| \leq \tfrac{1}{w}.
\end{align*}
Since the columns of $\X$ are non-decreasing, the $1$-norm sum
telescopes and we get $ \norm{\vec{\S\X\!-\!\X}}_1 \!\! = \!\!
\norm{\x_n}_{1} \!\! = \!\! \norm{\X}_{\infty}$.
\end{IEEEproof}

\begin{lemma}
\label{lem:shiftenergy}
Let $\X \in \Xc^{2L+w}$ have the decomposition
$\X = [\x_{-L}^{\t},\cdots,\x_{L+w-1}^{\t}]^{\t}$
and satisfy $[\X]_i = [\X]_{i_0}$ for $i>i_0$.
Then, $U(\S\X;\epsilon) - U(\X;\epsilon) \leq -U(\x_{i_{0}};\epsilon)$.
\end{lemma}
\begin{IEEEproof}
First, we rewrite the potential as the summation
\[ U(\X;\epsilon)=\!\!\!\sum_{i=-L}^{L+w-1}\!\!\g(\x_{i})\D\x_{i}^{\t}-G(\x_{i})-\!\sum_{i=-L}^{L}\!\! F\left(\left[\A\g(\X)\right]_{i};\epsilon\right).  \]
Next, we compute $U(\S\X;\epsilon)-U(\X;\epsilon)$
separately each of the three terms in $U(\X;\epsilon)$.
The first term, $T_1$, equals
\begin{align*}
T_{1} & =\!\sum_{i=-L+1}^{L+w-2}\!\g(\x_{i})\D\x_{i}^{\t}-\!\sum_{i=-L}^{L+w-1}\!\g(\x_{i})\D\x_{i}^{\t}\\
 & =-\g(\x_{L+w-1})\D\x_{L+w-1}^{\t}=-\g(\x_{i_0})\D\x_{i_0}^{\t}.
\end{align*}
Similarly, the second term is given by $T_2 = G(\x_{i_0})$.
For the third term, it follows from $F(\x;\epsilon)\geq 0$ and $[\A\g(\S\X)]_{i} =[\A\g(\X)]_{i-1}$ (for $i \in \mathcal{L}_{\g} \setminus \{-L\}$), that
\vspace{-1mm}
\[-F\left(\A\S\g(\X);\epsilon\right)
\leq - \sum_{i=-L}^{L-1}F\left([\A\g(\X)]_{i};\epsilon\right).  \]
Therefore, the third term satisfies
\vspace{-1mm}
\begin{align*}
T_{3} & =-F\left(\A\S\g(\X);\epsilon\right)+F\left(\A\g(\X);\epsilon\right)\\
 & \leq-\sum_{i=-L}^{L-1}F\left([\A\g(\X)]_{i};\epsilon\right)+\sum_{i=-L}^{L}F\left([\A\g(\X)]_{i};\epsilon\right)\\
 & =F\left([\A\g(\X)]_{L};\epsilon\right)=F\left(\g(\x_{i_0});\epsilon\right).
\end{align*}
Finally, we can write
\vspace{-1mm}
\begin{multline*}
U(\S\X;\epsilon)-U(\X;\epsilon) = T_1 + T_2 + T_3 \leq
\\ \g(\x_{i_0})\D\x_{i_0}^{\t}+G(\x_{i_0})
+F\left(\g(\x_{i_0});\epsilon\right)
 =-U(\x_{i_0};\epsilon).
\end{multline*}
\end{IEEEproof}

\begin{lemma}
For a fixed point of the one-sided SC system $\X$,
\begin{equation*}
\label{lemma:shift_differential}
\vec{U'(\X ; \epsilon)} \cdot \vec{\S \X - \X} = \bm{0}.
\end{equation*}
\end{lemma}

\begin{IEEEproof}
Rows $1,\ldots,i_{0}+L$ of $U'(\X;\epsilon)$ are zero since $\X$
is a fixed point of the one-sided spatially-coupled system. Also,
rows $L+i_{0}+1,\ldots,2L+(w-1)$ of $\S \X - \X$ are all-zero due to
the right boundary constraint. Hence, the inner product of these two
terms is identically zero.
\end{IEEEproof}

\begin{lemma}
\label{lem:xi0LimitExist}
Let $\X$ be a fixed point of the one-sided SC system. Then,
$\x^{\infty}(\x_{i_0};\epsilon)$ exists, $\x_{i_{0}} \preceq
\x^{\infty}(\x_{i_0};\epsilon)$, and $U(\x_{i_0};\epsilon) \geq U(\x^{\infty}(\x_{i_0};\epsilon);\epsilon)$.
\end{lemma}
\begin{IEEEproof}
From Lem.~\ref{lem:monotone}, $\X$ is non-decreasing.
From (\ref{eq:4}) and by the fact that $\X$ is a fixed point of the
one-sided SC system, it can be shown that
\begin{align*}
 \x_{i_{0}} &= \frac{1}{w} \sum_{k=0}^{w-1} \f \left(\frac{1}{w} \sum_{j=0}^{w-1} \g
 \left( \x_{i_{0}+j-k} \right); \epsilon \right) \\
  &\preceq \frac{1}{w} \sum_{k=0}^{w-1} \f \left(\frac{1}{w} \sum_{j=0}^{w-1} \g
 \left( \x_{i_{0}} \right); \epsilon \right) = \f \left( \g \left( \x_{i_{0}} \right); \epsilon \right).
\end{align*}
By applying Lem.~\ref{lem:LimitExists}, the claim of the lemma follows.
\end{IEEEproof}

\begin{lemma}
\label{lemma:hessian_bound}
  The norm of the Hessian, $U''(\X;\epsilon)$, of the SC potential is
  bounded by a constant independent of $L$ and $w$.
\end{lemma}
\begin{IEEEproof}
By direct computation, we obtain
\begin{align*}
  \norm{U''(\X;\epsilon)}_{\infty} &\leq \norm{\D}_{\infty} \big(g'_m + g''_m + 2\left(g_m'\right)^2f_m'\big)\triangleq K_{\f,\g},
\end{align*}
where $g'_m = \sup_{\x\in\Xc}\norm{\g'(\x)}_{\infty}$, $g''_m =
\sup_{\x\in\Xc}\norm{\g''(\x)}_{\infty}$ and $f'_m =
\sup_{\x\in\Xc}\norm{\f'(\x;\epsilon)}_{\infty}$.
\end{IEEEproof}

\begin{theorem}
\label{thm:main_theorem}
For a vector admissible system~$(\f,\g)$ with $\epsilon <
\epsilon^*$ and $w > (d K_{\f,\g})/( 2\Delta E(\epsilon))$, the only fixed point of the spatially-coupled system (Def.~\ref{eq:4}) is $\bm{0}$.
\end{theorem}
\begin{IEEEproof}
Fix $\epsilon < \epsilon^{*}$ and $w > (d K_{\f,\g}) / (2 \Delta E(\epsilon))$.
Suppose $\X \ne \bm{0}$ is the unique fixed point (Lem.~\ref{lem:monotone}) of the one-sided recursion
in Def.~\ref{def:OneSidedSC}.
Using Taylor's Theorem, the second-order expansion of $U( \S \X ; \epsilon)$ about $\X$ gives
\begin{align*}
&\tfrac{1}{2} \vec{\S \X - \X}^{\t} U''(\bm{Z}(t) ; \epsilon) \vec{\S \X - \X}  \\
&= U( \S \X; \epsilon) - U(\X ; \epsilon) - \vec{U'(\X ; \epsilon)} \cdot \vec{\S \X - \X} \\
&= U( \S \X; \epsilon) - U(\X ; \epsilon) \quad  \text{ ( Lem.~\ref{lemma:shift_differential} )} \\
& \le - U(\x_{i_{0}} ; \epsilon) \quad \quad \quad \quad \quad \ \text{ ( Lem.~\ref{lem:shiftenergy} )} \\
& \le - U\left( \x^{(\infty)} \left( \x_{i_{0}}; \epsilon \right) ;\epsilon \right) \ \,  \text{ ( Lem.~\ref{lem:xi0LimitExist} )}\\
& \le -\Delta E(\epsilon), \quad \quad \quad \quad \quad \quad
\text{ ( Def.~\ref{def:PotentialThreshold} ) }
\end{align*}
for some $t\in[0,1]$ with $\bm{Z}(t) = \X + t (\S \X - \X)$.
Taking the absolute value and using 
Lemmas~\ref{lem:shiftnormbound} and \ref{lemma:hessian_bound}
gives
\begin{align*}
\Delta E& (\epsilon)
\le \big\lvert \tfrac{1}{2} \vec{\S \X - \X}^{T} U''(\bm{Z}(t) ; \epsilon) \vec{\S \X - \X} \big\rvert \\
& \le \tfrac{1}{2}  \! \norm{\vec{\S \! \X  \! -  \! \X} }_{1} \! \norm{ U''(\bm{Z}(t) ;\epsilon) }_{\infty}  \! \norm{\vec{\S \X  \! -  \! \X} }_{\infty} \\
&\le \tfrac{1}{2} \norm{\X}_{\infty} K_{\f,\g} \tfrac{1}{w} \le \tfrac{d}{2w} K_{\f,\g}.
\end{align*}
This implies that $w \le (d K_{\f,\g})/\left(2 \Delta E (\epsilon) \right)$, but that contradicts the hypothesis.
Thus, the only fixed point for the one-sided spatially-coupled system is
the trivial fixed point $\bm{0}$. Also, the one-sided spatially-coupled
system upper bounds the two-sided system. Hence, the only fixed point of
the two-sided system is $\bm{0}$.
\end{IEEEproof}


\section{Applications}
\label{sec:applications}

In this section, we apply Theorem~\ref{thm:main_theorem} to a few coding problems that have vector DE recursions.
To save space, we rely on definitions and notation from~\cite{RU-2008,Nguyen-2012,Yedla-2012}.

\subsection{Potential Function and EBP EXIT Curves}
First, we introduce a connection between the
single-system potential function and the {\em EBP EXIT
curves} used in~\cite{Nguyen-isit12}.
As discussed in Rem.~\ref{rem:vecpara}, the potential function can be
constructed along a smooth and strictly increasing (w.r.t. the partial order)
path $\e(\epsilon)$ parameterized by $\epsilon \in \Ec$. Given a
parameter threshold $\epsilon'$ computed from the potential function, the
corresponding vector threshold $\e(\epsilon')$ is uniquely
determined.
%
%

Along the same monotone increasing curve $\e(\epsilon)$, the
$\x$-support $\Xc_f$ in Def.~\ref{def:eroot} can be defined.
Let $\x_{*} \triangleq \x^{(\infty)}({\bf 1},1)$ be the fixed point
when $\epsilon=1$. Then, a portion of the EBP EXIT curve can be
constructed along a smooth and strictly increasing (w.r.t. the
partial order) path, $\x(x) \in \Xc_f$, characterized by $x \in
[0,1]$ with $\x(1)=\x_{*}$.
Let $\epsilon(\x)$ be the $\epsilon$-set defined in
Def.~\ref{def:eroot}. It has been shown in
Lem.~\ref{lem:3properties} that $\epsilon(\x)$ is unique. If
$\epsilon(\x(x))$ is smooth in $x$, then the EBP EXIT
curve is given implicitly by $(\epsilon(\x(x)), \mathtt{h}^{\rm EBP}(\x(x)))$, where
$\mathtt{h}^{\rm EBP}(\x(x))$ is the associated EBP EXIT function
\cite{RU-2008,Nguyen-isit12}.

\begin{definition}
\label{def:TrialEntropy}
Given the EBP EXIT curve, the \emph{trial entropy} $P(\x)$ can be defined along the path $\x(x)$ with
\begin{align*}
 \label{def:TrialEntropy}
 P(\x(x)) &\triangleq P(\x_{*}) - \int_{x}^{1} \mathtt{h}^{\rm EBP}(\x(t)) \epsilon'(\x(t)) \cdot \x'(t) \mathrm{d}t.
\end{align*}
Note that the constant $P(\x_{*})$ in (\ref{def:TrialEntropy}) is
determined by the system under consideration. For example, in the
following applications, we know that $\x_{*} = \bm{1}$ and
$P(\x_{*})$ is equal to the design rate of the
system~\cite{Nguyen-isit12}. \iftrialPxI Moreover, let
$x^{\mathrm{Max}}$ be
\begin{align*}
 x^{\mathrm{Max}} \triangleq \min\{x\in[0,1] \mid P(\x(t))\geq0 \mathrm{\;for\;all\;} t\in
 [x,1]\}.
\end{align*}
The {\em Maxwell threshold} is denoted by $\epsilon^{\mathrm{Max}}$ and
defined by
\begin{align}
\label{eq:maxwell_thresh} \epsilon^{\mathrm{Max}} &= \min \left\{
\epsilon(\x(x)) \mid x \in \left[x^{\mathrm{Max}}, 1\right]
\right\}.
\end{align}
\fi
\end{definition}

\begin{remark}
The trial entropy is called \emph{path independent} if $P(\x_1(x_1))=P(\x_2(x_2))$ for
any two smooth paths $\x_1(x)$ and $\x_2(x)$ in $\Xc_f$ satisfying $\x_1(x_1)=\x_2(x_2)=\x$ for $x_1,x_2 \in[0,1]$.
This is required for a well-defined trial entropy $P(\x)$.
\end{remark}

\iftrialPxI
\begin{lemma}
\label{lem:Maxwell}

For a $P(\x)$ that is path independent, the definition of
$\epsilon^{\mathrm{Max}}$ can be rewritten as
\begin{align}
 \label{eq:maxwell_thresh2}
 \!\! \epsilon^{\mathrm{Max}} \triangleq \inf \left\{ \epsilon \in \Ec \, \big| \max_{\x \in \mathcal{F}(\epsilon)} \! P(\x)\!>\!0
 \text{ for all } \epsilon' \! \in \! [\epsilon,1] \right\}\!.\!
\end{align}

\end{lemma}
\begin{IEEEproof}[Not complete proof.]
Let $\x(x) \in \Xc_f$ be a smooth and monotone increasing path such
that $\epsilon(\x(x))$ is smooth in $x$. Let
$\epsilon^{\mathrm{Max}}$ be as defined in
Def.~\ref{def:TrialEntropy}. By the smoothness of $\epsilon(\x(x))$
and the fact that $\epsilon(\x(1))=\epsilon({\bf 1})=1$, one can see
that $\{\epsilon(\x(x)) \mid x\in[x^{\mathrm{Max}},1]\} =
[\epsilon^{\mathrm{Max}}, 1]$ and $\x(x) \in
\mathcal{F}(\epsilon(\x(x)))$. Since $P(\x)$ is path independent,
Def.~\ref{def:TrialEntropy} implies that, for all $\epsilon \geq
\epsilon^{\mathrm{Max}}$, there is some $\x \in
\mathcal{F}(\epsilon)$ such that $P(\x)>0$.
%
\end{IEEEproof}
\else
\begin{definition}[{c.f. \cite[Conj.~1]{Measson-it08}}]
\label{def:Maxwell}

For a $P(\x)$ that is path independent, the {\em Maxwell Threshold}
$\epsilon^{\mathrm{Max}}$ is defined by
\begin{equation}
 \label{eq:maxwell_thresh}
 \epsilon^{\mathrm{Max}}\! =\! \inf \left\{ \epsilon \in \Ec \, \big| \max_{\x \in \mathcal{F}(\epsilon)} \! P(\x)\!>\!0
 \right\}\!.\!\!
\end{equation}

\end{definition}

\begin{remark}
In \cite[Conj.~1]{Measson-it08}, the Maxwell threshold is defined implicitly as the conjectured MAP threshold for irregular LDPC codes on the BEC.
When the MAP threshold is determined by stability, the definition in \cite[Conj.~1]{Measson-it08}, however, does not identify the stability threshold correctly.
This can be repaired simply by replacing the interval $(0,1]$ in their conjecture with the interval $[0,1]$ and defining the $\epsilon$ associated with the fixed point $x=0$ to be the limit as $x \to 0$.
For irregular LDPC codes, Def.~\ref{def:Maxwell} is equivalent to the corrected definition in \cite[Conj.~1]{Measson-it08}.
We also believe that it is the correct extension for our more general setup.

\end{remark}

\fi

\vspace{-1mm}
\subsection{Noisy Slepian-Wolf Problem with Erasures }
\label{sec:erasure-slepian-wolf}
\vspace{-0.3mm}

Two correlated discrete memoryless sources are encoded by two
independent linear encoding functions, which are then transmitted
through two independent erasure channels with erasure rates
$\epsilon_1$ and $\epsilon_2$, respectively.  We consider an erasure
correlation model between the two sources. More specifically, let
$Z$ be a Bernoulli-$p$ random variable such that the two sources are
the same Bernoulli-$\frac{1}{2}$ random variable if $Z=1$ and are
i.i.d. Bernoulli-$\frac{1}{2}$ random variables if $Z=0$. The
decoder is assumed to have access to the side information $Z$.

Assume that the $i$-th source sequence is mapped into an LDPC code with
degree distribution (d.d.) $(\lambda_i,\rho_i)$ and design rate
$\gamma=1-L_i'(1)/R_i'(1)$ using a punctured
systematic encoder of rate $\gamma/(1-\gamma)$.  The fraction of
punctured systematic bits is $\gamma$ (see~\cite{Yedla-isit11} for details).
Let $x^{(\ell)}_{1}$ (resp. $x^{(\ell)}_{2}$) be
the average erasure rate of messages, from bit nodes to check nodes,
corresponding to source $1$ (resp. $2$) and $\x^{(\ell)} =
[x^{(\ell)}_1,x^{(\ell)}_2]$. Let $\mathcal{C}:[0,1]\to[0,1]^2$,
$\epsilon\mapsto [\epsilon_1(\epsilon), \epsilon_2(\epsilon)]$ with
$[\epsilon_1(0),\epsilon_2(0)] = \bm{0}$, be continuous and
monotonically increasing. The DE recursion in~\cite{Yedla-isit11} is
easily generalized to asymmetric d.d.s and can be written in the form of~(\ref{eq:1}) with
\vspace{-0.2mm}
\begin{align*}
  \psi(x;\epsilon) &\triangleq (1-\gamma)\epsilon + \gamma (1-p + px),\\
  \f(\x;\epsilon) &\triangleq [\psi(L_2(x_2);\epsilon_1(\epsilon))\lambda_1(x_1), \psi(L_1(x_1);\epsilon_2(\epsilon)) \lambda_2(x_2)],\\
  \g(\x) &\triangleq [1 - \rho_1(1-x_1), 1 - \rho_2(1-x_2)].
\end{align*}
Using Def.~\ref{def:potential} and $\D = \text{diag}\left(L_1'(1),L_2'(1)\right)$, one finds that
\begin{align}
 F(\x; \epsilon) &= \psi(L_{1}(x_{1}); \epsilon_{2}(\epsilon)) L_{2}(x_{2}) \nonumber \\
&\quad + \psi(L_{2}(x_{2}); \epsilon_{1}(\epsilon)) L_{1}(x_{1}) -  \gamma p  L_{1}(x_{1}) L_{2}(x_{2}), \nonumber \\
 G(\x) &= \sum\nolimits_{k=1}^{2} L_{k}'(1) \Big( x_{k} + \frac{R_{k}(1-x_{k}) - 1}{R_{k}'(1)} \Big). \nonumber
\end{align}

For asymmetric d.d.s, one can also generalize both the trial entropy $P(\x)$, from~\cite[Lem.~4]{Nguyen-isit12}, and the mapping $\e(\x) = [\epsilon^{[1]}(\x), \epsilon^{[2]}(\x)]$, from~\cite[Sec.~II-A]{Nguyen-isit12}.
\ifmaxwell
  This gives
\begin{align*}
  U(\x;\epsilon) = (1-\gamma)\left( (\e(\x)\!-\![\epsilon_1(\epsilon),\epsilon_2(\epsilon)])\bm{L}(\g(\x))^{\intercal}\!
  - \! P(\x)\right)\!,
\end{align*}
where $\bm{L}(\g(\x)) = [L_1(g_1(x_1)),L_2(g_2(x_2))]$.
Since $Q(\x) \triangleq U(\x;\epsilon(\x))$, substituting 
$\epsilon \mapsto \epsilon(\x)$ into $U(\x;\epsilon)$ shows that
\begin{align}
 \label{eq:SWEPQ}
 Q(\x) &= -(1-\gamma)P(\x).
\end{align}

\iffalse
Therefore, we find that the Maxwell threshold from
Def.~\ref{def:Maxwell} is equivalent to the standard Maxwell
threshold~\cite[Conj.~1]{Measson-it08}. \else
\fi

\begin{lemma}
\label{lem:esw_ldpc}
Let $\epsilon^*$ be the potential threshold from
(\ref{eq:PotentialInQ}). Then, $\epsilon^* = \epsilon^{\text{Max}}$,
where $\epsilon^{\text{Max}}$ is the Maxwell threshold defined by
(\ref{eq:maxwell_thresh}).
\end{lemma}
\begin{IEEEproof}
 By substituting (\ref{eq:SWEPQ}) into (\ref{eq:maxwell_thresh}), it can be shown that
 \begin{align}
  \label{eq:SWEMax}
  \epsilon^{\mathrm{Max}} \triangleq \inf \left\{ \epsilon \! \in \! \Ec \, \big| \min_{\x \in \mathcal{F}(\epsilon)} \! Q(\x) \! < \! 0 \right\}.\!\!
 \end{align}
 Since the condition in (\ref{eq:PotentialInQ}) is the complement of the condition in (\ref{eq:SWEMax}) and the minimum $\min_{\x \in \mathcal{F}(\epsilon)} Q(\x)$ is strictly decreasing in $\epsilon$, these thresholds are equal.
\end{IEEEproof}
\fi

\begin{corollary}
  Applying Theorem~\ref{thm:main_theorem} shows that, if $\epsilon <
  \epsilon^{\text{Max}}$ and $w >  K_{\f,\g} / \Delta E(\epsilon)$, then
  the SC Slepian-Wolf DE recursion must converge to the zero matrix.
\end{corollary}

\begin{remark}
  For special cases, one can use the methods in~\cite{Nguyen-isit12,Nguyen-2012,Yedla-2012} to show that the Maxwell threshold defined above is an upper bound on the MAP threshold.
  These references also show that, for regular LDPC codes with fixed rate and increasing degrees, the upper bound approaches the information-theoretic limit.
  Therefore, SC regular LDPC codes are universal (e.g., see~\cite{Yedla-isit11}) for this problem.
\end{remark}

\ifmaxwell
\else
\begin{remark}
  For a fixed ensemble, one can compute the region achievable via
  spatial-coupling by considering curves of
  the form $\epsilon\mapsto [\epsilon,\theta\epsilon]$, for
  $\theta\in[0,\infty)$. We note that, for regular codes with fixed rate,
  this region approaches to the Slepian-Wolf region (e.g., see~\cite{Yedla-isit11}) as the degrees are
  increased. Therefore, SC regular LDPC codes are
  universal (with respect to channel parameters) for this problem.
\end{remark}
\fi


\vspace{-0.5mm}

\subsection{Erasure Multiple-Access Channel}
\label{sec:eras-mult-access}
\vspace{-0.3mm}

We consider the two-user MAC channel with erasure
noise (EMAC) from~\cite{Kudekar-isit11-MAC}. Let the
inputs be $X^{[1]},X^{[2]}\in\{\pm1\}$ and the output be
\begin{align*}
Y=\begin{cases}
X^{[1]}+X^{[2]} & \text{with probability\,\,}1-\epsilon,\\
? & \text{with probability\,\,}\epsilon
\end{cases}.
\end{align*}
\vspace{-2mm}

Assume that the source sequences are encoded by LDPC codes with
d.d.s $(\lambda_1,\rho_1)$ and $(\lambda_2,\rho_2)$.  Let $x^{(\ell)}_{1}$
(resp. $x^{(\ell)}_{2}$) be the average erasure rate of messages from
bit nodes to check nodes corresponding to user $1$ (resp. $2$) and
$\x^{(\ell)} = [x^{(\ell)}_1,x^{(\ell)}_2]$. In~\cite{Nguyen-isit12},
the DE recursion is written as~(\ref{eq:1}), with
\vspace{-1mm}
\begin{align*}
  \psi(x;\epsilon) &\triangleq \epsilon + (1-\epsilon)x/2,\\
  \f(\x;\epsilon) &\triangleq [\psi(L_2(x_2);\epsilon)\lambda_1(x_1), \psi(L_1(x_1);\epsilon) \lambda_2(x_2)],\\
  \g(\x) &\triangleq [1 - \rho_1(1-x_1), 1 - \rho_2(1-x_2)].
\end{align*}
Using Def.~\ref{def:potential} and $\bm{D} = \text{diag}\left(L_1'(1),L_2'(1)\right)$, one finds that
\vspace{-1mm}
\begin{align*}
  F(\x;\epsilon) & =\epsilon [ L_{1}(x_{1}) + L_{2}(x_{2}) ] + (1- \epsilon) L_{1}(x_{1}) L_{2}(x_{2}) /2, \\
  G(\x)  &= \sum\nolimits_{k=1}^{2} L_{k}'(1) \Big( x_{k} + \frac{R_{k}(1-x_{k}) - 1}{R_{k}'(1)} \Big).
\end{align*}
Let the trial entropy, $P(\x)$, and the $\epsilon(\x)$ be defined by~\cite[Lem.~10]{Nguyen-isit12}.
Then, $U(\x;\epsilon)$ equals
\vspace{-1mm}
\begin{align*}
( \epsilon(\x) - \epsilon) [\bm{L}(\g(x))\bm{1}^\t - \tfrac{1}{2} L_{1}(g_{1}(x_{1})) L_{2}(g_{2}(x_{2}))] - P(\x),
\end{align*}
\ifmaxwell
and substituting $\epsilon \mapsto \epsilon(\x)$ implies that $Q(\x) = - P(\x)$.
Similar to Lemma~\ref{lem:esw_ldpc}, it can also be shown that
$\epsilon^* = \epsilon^\mathrm{Max}$.
\else
\begin{lemma}
\label{lem:esw_ldpc}
Let $\epsilon^*$ be the potential threshold from (\ref{eq:3}).
Then, $\epsilon^* = \epsilon^{\text{Max}}$, where $\epsilon^{\text{Max}}$ is the Maxwell threshold defined by
  \begin{align}
    \label{eq:esw6}
    \epsilon^{\text{Max}} = \min \left\{\epsilon \in [0,1] \mid \exists \x \in \mathcal{F}(\epsilon) \; \mathrm{s.t.} \; P(\x)=0\right\},
  \end{align}
  where $\mathcal{F}(\epsilon) = \{ \x \in \Xc \mid \f(\g(\x);\epsilon)=\x \}$.
\end{lemma}
\begin{IEEEproof}
  Any fixed-point $\x \in \mathcal{F}(\epsilon)$ is supported by a unique channel parameter $\epsilon(\x)$.
  Therefore, $U(\x;\epsilon) = -2P(\x)/(1-\epsilon)$ for all $\x \in \mathcal{F}(\epsilon)$.
  This implies that $U(\x^{\mathrm{Max}};\epsilon^{\mathrm{Max}}) =  -2P(\x^{\mathrm{Max}})/(1-\epsilon) = 0$.
  From Def.~\ref{def:PotentialThreshold}, we know $\epsilon^*\leq
  \epsilon^{\text{Max}}$.  It can also be shown that $P(\x^*)=0$,
  and thus, $\epsilon^{\text{Max}}\leq \epsilon^*$. This implies the equality.
\end{IEEEproof}
\fi

\begin{corollary}
  Applying Theorem~\ref{thm:main_theorem} shows that, if $\epsilon <
  \epsilon^{\text{Max}}$ and $w > K_{\f,\g}/\Delta E(\epsilon)$, then
  the SC DE recursion for the erasure MAC channel must converge to the zero matrix.
\end{corollary}

\begin{remark}
  For special cases, one can use the methods in~\cite{Nguyen-isit12,Nguyen-2012,Yedla-2012} to show that the Maxwell threshold defined above is an upper bound on the MAP threshold.
  These references also show that, for regular LDPC codes with fixed rate and increasing degrees, the upper bound approaches the information-theoretic limit of the EMAC channel.
\end{remark}


\vspace{-0.5mm}
\subsection{General Protograph Codes on the BEC}
\label{sec:protograph}
\vspace{-0,3mm}

Consider the protograph ensemble~\cite{Thorpe-ipn03} defined by an $m\times n$ protograph parity-check matrix $H$ (e.g., $H=[3\;3]$ defines a (3,6)-regular code) and let $[k]$ denote the set $\{1,2,\ldots,k\}$.
Let the dimension of the recursion, $d$, equal the number of non-zero entries in $H$ and let the functions $r:[d]\to[m]$, $c:[d]\to[n]$, and $e:[d]\to\left\{ 1,2,\ldots\right\} $ map the index of each non-zero entry to its row, column, and value (i.e., $e(k) = H_{r(k),c(k)}$ for $k\in[d]$).
Let $\epsilon_{j}(\epsilon)$ be the erasure probability of the $j$-th bit node in the protograph as a function of the channel parameter $\epsilon$.
Then, the bit- and check-node DE update functions $\f(\x),\g(\x)$ are given by
\iffalse
\begin{equation*}
f_{k}\!(\x;\epsilon) \!\! = \! \epsilon_{c(k)}\!(\epsilon) \hspace{-7mm} \prod_{i\in[d]:c(i)=c(k)} \hspace{-6.5mm} x_{i}^{e(i)-\delta_{i,k}},
\hspace{0.5mm}
\!g_{k}\!(\x) \!\! = \!\! 1 \,\! - \hspace{-6.75mm} \prod_{j\in[d]:r(j)=r(k)} \hspace{-7mm} (1 - x_{j})^{e(j)-\delta_{i,k}}\!,
\end{equation*}
\else
\begin{align*}
f_{k} (\x;\epsilon) &= \epsilon_{c(k)}(\epsilon) \hspace{-3mm} \prod_{i\in[d]:c(i)=c(k)} \hspace{-2.5mm} x_{i}^{e(i)-\delta_{i,k}} \\
g_{k}(\x) &= 1 - \hspace{-3.75mm} \prod_{j\in[d]:r(j)=r(k)} \hspace{-3mm} (1 - x_{j})^{e(j)-\delta_{i,k}},
\end{align*}
where $\delta_{i,j}$ is the Kronecker delta function. From this, we
make an educated guess that the bit- and check-node potentials are
\begin{align*}
F(\x;\epsilon) &= \sum\nolimits_{j=1}^{n}\epsilon_{j}(\epsilon)\prod\nolimits_{i\in[d]:c(i)=j}x_{i}^{e(i)}\\
G(\x) &= \sum_{k=1}^{d} e(k)x_{k} - \sum_{i=1}^{m} \left[ 1 - \prod_{j\in[d]:r(j)=i}(1-x_{j})^{e(j)} \right].
\end{align*}
Since each non-zero entry in $H$ appears in only one row and one
column, it is easy to verify that \vspace{-2mm}
\begin{align*}
\frac{\mathrm{d}}{\mathrm{d}x_{k}}F(\x;\epsilon) = e(k)f_{k}(\x;\epsilon) \;\; \mathrm{and} \;\;
\frac{\mathrm{d}}{\mathrm{d}x_{k}}G(\x) = e(k)g_{k}(\x).
\end{align*}
This shows that one can choose $\D=\mathrm{diag}\left(e(1),\ldots,e(d)\right)$ and then apply Def.~\ref{def:potential} to define a potential function for the protograph DE update.
It is easy to verify that the DE equations comprise a vector admissible system.
Therefore, we conjecture that the fixed-point potential, $Q(\x)$, will also be a scalar multiple of the trial entropy defined by integration of the BP EXIT curve~\cite{RU-2008}.

\vspace{-0.5mm}
\section{Conclusions}

Based on the work in~\cite{Yedla-istc12}, a new theorem is presented
that provides a simple proof of threshold saturation for a broad class of vector
recursions.  The conditions of the theorem are verified for the
density-evolution equations associated with: (i) irregular LDPC codes
for a Slepian-Wolf problem with erasures, (ii) irregular LDPC codes
on the erasure multiple-access channel, and (iii) protograph
codes on BEC.  This provides the first proof of threshold saturation for
these systems. Along with the results in \cite{Nguyen-isit12,Nguyen-2012,Yedla-2012},
this also shows that SC codes are universal (e.g., see~\cite{Yedla-isit11}) for the noisy Slepian-Wolf problem
with erasures.

\subsubsection*{Acknowledgment}
The authors thank Kenta Kasai for identifying a few subtle errors in an earlier version of this manuscript and Andrew J.\ Young for suggestions on and assistance with the final version.

\bibliographystyle{ieeetr}
\bibliography{WCLabrv,WCLbib,WCLnewbib}

\begin{thebibliography}{10}

\bibitem{Felstrom-it99}
J.~Felstrom and K.~S. Zigangirov, ``Time-varying periodic convolutional codes
  with low-density parity-check matrix,'' {\em IEEE Trans.\ Inform.\ Theory},
  vol.~45, no.~6, pp.~2181--2191, 1999.

\bibitem{Sridharan-aller04}
A.~Sridharan, M.~Lentmaier, D.~J. Costello, and K.~S. Zigangirov, ``Convergence
  analysis of a class of {LDPC} convolutional codes for the erasure channel,''
  in {\em Proc.\ Annual Allerton Conf.\ on Commun., Control, and Comp.},
  (Monticello, IL), pp.~953--962, 2004.

\bibitem{Lentmaier-isit05}
M.~Lentmaier, A.~Sridharan, K.~S. Zigangirov, and D.~J. Costello, ``Terminated
  {LDPC} convolutional codes with thresholds close to capacity,'' in {\em
  Proc.\ IEEE Int.\ Symp.\ Inform.\ Theory}, (Adelaide, Australia),
  pp.~1372--1376, 2005.

\bibitem{Lentmaier-it10}
M.~Lentmaier, A.~Sridharan, D.~J. Costello, and K.~S. Zigangirov, ``Iterative
  decoding threshold analysis for {LDPC} convolutional codes,'' {\em IEEE
  Trans.\ Inform.\ Theory}, vol.~56, pp.~5274--5289, Oct. 2010.

\bibitem{Kudekar-istc10}
S.~Kudekar, C.~M\'easson, T.~Richardson, and R.~Urbanke, ``Threshold saturation
  on {BMS} channels via spatial coupling,'' in {\em Proc.\ Int.\ Symp.\ on
  Turbo Codes \& Iterative Inform.\ Proc.}, pp.~309--313, Sept. 2010.

\bibitem{Rathi-isit11}
V.~Rathi, R.~Urbanke, M.~Andersson, and M.~Skoglund, ``Rate-equivocation
  optimally spatially coupled {LDPC} codes for the {BEC} wiretap channel,'' in
  {\em Proc.\ IEEE Int.\ Symp.\ Inform.\ Theory}, (St.\ Petersburg, Russia),
  pp.~2393--2397, July 2011.

\bibitem{Yedla-isit11}
A.~Yedla, H.~D. Pfister, and K.~R. Narayanan, ``Universality for the noisy
  {S}lepian-{W}olf problem via spatial coupling,'' in {\em Proc.\ IEEE Int.\
  Symp.\ Inform.\ Theory}, (St.\ Petersburg, Russia), pp.~2567--2571, July
  2011.

\bibitem{Kudekar-isit11-DEC}
S.~Kudekar and K.~Kasai, ``Threshold saturation on channels with memory via
  spatial coupling,'' in {\em Proc.\ IEEE Int.\ Symp.\ Inform.\ Theory}, (St.\
  Petersburg, Russia), pp.~2562--2566, July 2011.

\bibitem{Nguyen-arxiv11}
P.~S. Nguyen, A.~Yedla, H.~D. Pfister, and K.~R. Narayanan, ``Spatially-coupled
  codes and threshold saturation on intersymbol-interference channels.'' to be
  submitted to {\em IEEE Trans. on Inform. Theory}, [Online]. Available:
  http://arxiv.org/abs/1107.3253, 2012.

\bibitem{Nguyen-icc12}
P.~S. Nguyen, A.~Yedla, H.~D. Pfister, and K.~R. Narayanan, ``Threshold
  saturation of spatially-coupled codes on intersymbol-interference channels,''
  in {\em Proc.\ IEEE Int.\ Conf.\ Commun.}, (Ottawa, Canada), pp.~2209--2214,
  June 2012.

\bibitem{Kudekar-arxiv12}
S.~Kudekar, T.~Richardson, and R.~Urbanke, ``Spatially coupled ensembles
  universally achieve capacity under belief propagation.'' Arxiv preprint
  arXiv:1201.2999, 2012.

\bibitem{Kudekar-it11}
S.~Kudekar, T.~J. Richardson, and R.~L. Urbanke, ``Threshold saturation via
  spatial coupling: {W}hy convolutional {LDPC} ensembles perform so well over
  the {BEC},'' {\em IEEE Trans.\ Inform.\ Theory}, vol.~57, no.~2,
  pp.~803--834, 2011.

\bibitem{Lentmaier-isit10}
M.~Lentmaier and G.~P. Fettweis, ``On the thresholds of generalized {LDPC}
  convolutional codes based on protographs,'' in {\em Proc.\ IEEE Int.\ Symp.\
  Inform.\ Theory}, (Austin, TX), pp.~709--713, 2010.

\bibitem{Hassani-itw10}
S.~H. Hassani, N.~Macris, and R.~Urbanke, ``Coupled graphical models and their
  thresholds,'' in {\em Proc.\ IEEE Inform.\ Theory Workshop}, (Dublin,
  Ireland), pp.~1--5, 2010.

\bibitem{Hassani-jsm12}
S.~H. Hassani, N.~Macris, and R.~Urbanke, ``Chains of mean-field models,'' {\em
  J. Stat. Mech.}, p.~P02011, 2012.

\bibitem{Takeuchi-isit11}
K.~Takeuchi, T.~Tanaka, and T.~Kawabata, ``Improvement of {BP}-based {CDMA}
  multiuser detection by spatial coupling,'' in {\em Proc.\ IEEE Int.\ Symp.\
  Inform.\ Theory}, (St.\ Petersburg, Russia), pp.~1489--1493, July 2011.

\bibitem{Schlegel-isit11}
C.~Schlegel and D.~Truhachev, ``Multiple access demodulation in the lifted
  signal graph with spatial coupling,'' in {\em Proc.\ IEEE Int.\ Symp.\
  Inform.\ Theory}, (St.\ Petersburg, Russia), pp.~2989--2993, July 2011.

\bibitem{Jian-isit12}
Y.-Y. Jian, H.~D. Pfister, and K.~R. Narayanan, ``Approaching capacity at high
  rates with iterative hard-decision decoding,'' in {\em Proc.\ IEEE Int.\
  Symp.\ Inform.\ Theory}, pp.~2696--2700, 2012.

\bibitem{Kudekar-aller10}
S.~Kudekar and H.~D. Pfister, ``The effect of spatial coupling on compressive
  sensing,'' in {\em Proc.\ Annual Allerton Conf.\ on Commun., Control, and
  Comp.}, (Monticello, IL), pp.~347--353, Oct. 2010.

\bibitem{Krzakala-arxiv11}
F.~Krzakala, M.~M{\'e}zard, F.~Sausset, Y.~Sun, and L.~Zdeborov{\'a},
  ``Statistical physics-based reconstruction in compressed sensing.'' Arxiv
  preprint arXiv:1109.4424, Sept. 2011.

\bibitem{Donoho-arxiv11}
D.~L. Donoho, A.~Javanmard, and A.~Montanari, ``Information-theoretically
  optimal compressed sensing via spatial coupling and approximate message
  passing.'' Arxiv preprint arXiv:1112.0708, Dec. 2011.

\bibitem{Yedla-istc12}
A.~Yedla, Y.-Y. Jian, P.~S. Nguyen, and H.~D. Pfister, ``A simple proof of
  threshold saturation for coupled scalar recursions,'' in {\em Proc.\ Int.\
  Symp.\ on Turbo Codes \& Iterative Inform.\ Proc.}, 2012.
\newblock Arxiv preprint arXiv:1204.5703, 2012.

\bibitem{Takeuchi-arxiv11}
K.~Takeuchi, T.~Tanaka, and T.~Kawabata, ``A phenomenological study on
  threshold improvement via spatial coupling.'' Arxiv preprint arXiv:1102.3056,
  2011.

\bibitem{Kudekar-unpub12}
S.~Kudekar, T.~Richardson, and R.~Urbanke, ``Wave-like solutions of general
  one-dimensional spatially coupled systems.'' submitted to {\em IEEE Trans. on
  Inform. Theory}, [Online]. Available: http://arxiv.org/abs/1208.5273, Aug.
  2012.

\bibitem{Kudekar-isit11-MAC}
S.~Kudekar and K.~Kasai, ``Spatially coupled codes over the multiple access
  channel,'' in {\em Proc.\ IEEE Int.\ Symp.\ Inform.\ Theory}, (St.\
  Petersburg, Russia), pp.~2816--2820, July 2011.

\bibitem{Nguyen-isit12}
P.~S. Nguyen, A.~Yedla, H.~D. Pfister, and K.~R. Narayanan, ``On the maximum a
  posteriori decoding thresholds of multiuser systems with erasures,'' in {\em
  Proc.\ IEEE Int.\ Symp.\ Inform.\ Theory}, (Cambridge, MA), pp.~2711--2715,
  July 2012.

\bibitem{Magnus-1999}
J.~R. Magnus and H.~Neudecker, {\em Matrix Differential Calculus With
  Applications in Statistics and Econometrics}.
\newblock Wiley Series in Probability and Statistics, John Wiley, 1999.

\bibitem{Vontobel-acorn09}
P.~Vontobel, ``Message-passing iterative decoding and linear programming
  decoding: news and views.'' Lectures at 2009 {ACoRN} Spring School,
  University of South Australia, Adelaide, Australia. [Available online at
  www.pseudocodewords.info], Nov. 2009.

\bibitem{Walsh-com10}
J.~M. Walsh and P.~A. Regalia, ``On the relationship between belief propagation
  decoding and joint maximum likelihood detection,'' {\em IEEE Trans.\
  Commun.}, vol.~58, no.~10, pp.~2753--2758, 2010.

\bibitem{RU-2008}
T.~J. Richardson and R.~L. Urbanke, {\em Modern Coding Theory}.
\newblock Cambridge University Press, 2008.

\bibitem{Nguyen-2012}
P.~S. Nguyen, {\em Advanced Coding Techniques with Applications to Storage
  Systems}.
\newblock PhD thesis, Texas A\&M University, College Station, TX, 2012.

\bibitem{Yedla-2012}
A.~Yedla, {\em Universality for Multi-Terminal Problems via Spatial Coupling}.
\newblock PhD thesis, Texas A\&M University, College Station, TX, 2012.

\bibitem{Measson-it08}
C.~M{\'e}asson, A.~Montanari, and R.~L. Urbanke, ``Maxwell construction: The
  hidden bridge between iterative and maximum a posteriori decoding,'' {\em
  IEEE Trans.\ Inform.\ Theory}, vol.~54, pp.~5277--5307, Dec. 2008.

\bibitem{Thorpe-ipn03}
J.~Thorpe, ``Low-density parity-check ({LDPC}) codes constructed from
  protographs,'' {\em IPN Progress Report}, vol.~42, Aug. 2003.

\end{thebibliography}

\end{document}